\documentclass[twocolumn,journal]{IEEEtran}

\usepackage{amsfonts}
\usepackage{amsmath}
\usepackage{amsthm}
\usepackage{amssymb}
\usepackage{graphicx}
\usepackage[T1]{fontenc}
\usepackage{supertabular}
\usepackage{longtable}
\usepackage[usenames,dvipsnames]{color}
\usepackage{bbm}
\usepackage{fancyhdr}
\usepackage{breqn}
\usepackage{fixltx2e}
\usepackage{capt-of}
\usepackage{tikz}
\usetikzlibrary{matrix}
\usepackage{endnotes}
\usepackage{soul}
\usepackage{marginnote}

\newcommand{\mathsym}[1]{}
\newcommand{\unicode}[1]{}

\usepackage{hyperref}
\usepackage{caption}
\usepackage{graphicx}
\graphicspath{{figures/}{/Users/nntaleb/Dropbox/Latex/SilentRisk/}}
 
\usepackage[framemethod=TikZ]{mdframed}
\usepackage[framemethod=TikZ]{mdframed}
\usepackage{framed}

\usepackage{enumitem}

\graphicspath{{figures/}{/Users/nntaleb/Dropbox/Latex/Collectednew/}}
\mdfsetup{%
nobreak=true,
   middlelinecolor=black,
   middlelinewidth=1pt,
   backgroundcolor=yellow!20,
   roundcorner=3pt}

\mdtheorem[nobreak=true,outerlinewidth=1,
backgroundcolor=yellow!50, outerlinecolor=black,innertopmargin=0pt,splittopskip=\topskip,skipbelow=\baselineskip, skipabove=\baselineskip,ntheorem,roundcorner=5pt,font=\itshape]{theorem}{Theorem}

\mdtheorem[nobreak=true,outerlinewidth=1,
backgroundcolor=gray!10, outerlinecolor=black,innertopmargin=0pt,splittopskip=\topskip,skipbelow=\baselineskip, skipabove=\baselineskip,ntheorem,roundcorner=5pt,font=\itshape]{remark}{Rule}

\mdtheorem[nobreak=true,outerlinewidth=1,
backgroundcolor=pink!30, outerlinecolor=black,innertopmargin=0pt,splittopskip=\topskip,skipbelow=\baselineskip, skipabove=\baselineskip,ntheorem,roundcorner=5pt,font=\itshape]{quaestio}{Quaestio}

\mdtheorem[nobreak=true,outerlinewidth=1,
backgroundcolor=yellow!50, outerlinecolor=black,innertopmargin=5pt,splittopskip=\topskip,skipbelow=\baselineskip, skipabove=\baselineskip,ntheorem,roundcorner=5pt,font=\itshape]{background}{Background}

\usepackage{hyperref}
\title{\color{Brown} 
Tail Risk of Contagious Diseases 
}

\author{
    \IEEEauthorblockN{ Pasquale Cirillo\IEEEauthorrefmark{1} and Nassim Nicholas Taleb\IEEEauthorrefmark{2} }\\
    
    \IEEEauthorblockA{\IEEEauthorrefmark{1}Applied Probability Group, Delft University of Technology}
    
    \IEEEauthorblockA{\IEEEauthorrefmark{2}Tandon School of Engineering, New York University}
    
    \thanks{April 18, 2020. Corresponding author: Nassim Nicholas Taleb: nnt1@nyu.edu}
    
Forthcoming, \textit{Nature Physics}}

\begin{document}

\maketitle 

\begin{mdframed}
\begin{abstract} 
Applying a modification of Extreme value Theory (thanks to a dual distribution technique by the authors in \cite{CiTa}) on data over the past 2,500 years, we show that pandemics are extremely fat-tailed in terms of fatalities, with a marked potentially existential risk for humanity. 

Such a macro property should invite the use of Extreme Value Theory (EVT) rather than naive interpolations and expected averages for risk management purposes. An implication is that potential tail risk overrides conclusions on decisions derived from compartmental epidemiological models and similar approaches.
\end{abstract} 
\end{mdframed}

\section{Introduction and Policy Implications}

We examine the distribution of fatalities from major pandemics in history (spanning about 2,500 years), and build a statistical picture of their tail properties. Using tools from Extreme Value Theory (EVT), we show for 
that the distribution of the victims of infectious diseases is extremely fat-tailed, more than what one could be led to believe from the outset\footnote{In this comment we do not discuss the possible generating mechanisms behind these fat tails, a topic of separate research. Networks analysis, e.g. \cite{AlbertBarabasi}, proposes mechanisms for the spreading of contagion and the existence of super spreaders, a plausible joint cause of fat tails. Likewise simple automata processes can lead to high uncertainty of outcomes owing to ``computational irreducibility" \cite{Wolfram}.}.

A non-negative continuous random variable $X$ is fat-tailed, in the regular variation class, if its survival function $S(x)=P(X\geq x)$ decays as a power law $x^{-\frac{1}{\xi}}$, the more we move into the tail\footnote{More technically, a non-negative continuous random variable $X$ has a fat-tailed distribution (in the maximum domain of attraction of the Fr\'echet distribution), if its survival function is regularly varying, i.e. $S(x)=L(x) x^{-\frac{1}{\xi}}$, where $L(x)$ is a slowly varying function, such that $\lim_{x \to \infty}\frac{L(cx)}{L(x)}=1$ for $c>0$ \cite{deHaan,Embrechts}.}, that is for $x$ growing towards the right endpoint of $X$. The parameter $\xi$ is known as the tail parameter, and it governs the fatness of the tail (the larger $\xi$ the fatter the tail) and the existence of moments ($E[X^p]<\infty$ if and only if $\xi < 1/p$). In some literature, e.g. \cite{Clauset}, the tail index is re-parametrized as $\alpha=1/\xi$, and its interpretation is naturally reversed.

While it is known that fat tails represent a common--yet often ignored \cite{TaStat} in modeling--regularity in many fields of science and knowledge \cite{Clauset}, for the best of our knowledge, only war casualties and operational risk losses show a behavior \cite{CiTa,CiTaQF,Neslehova} as erratic and wild as the one we observe for pandemic fatalities.

\begin{figure} 
\centering
\includegraphics[width=\linewidth]{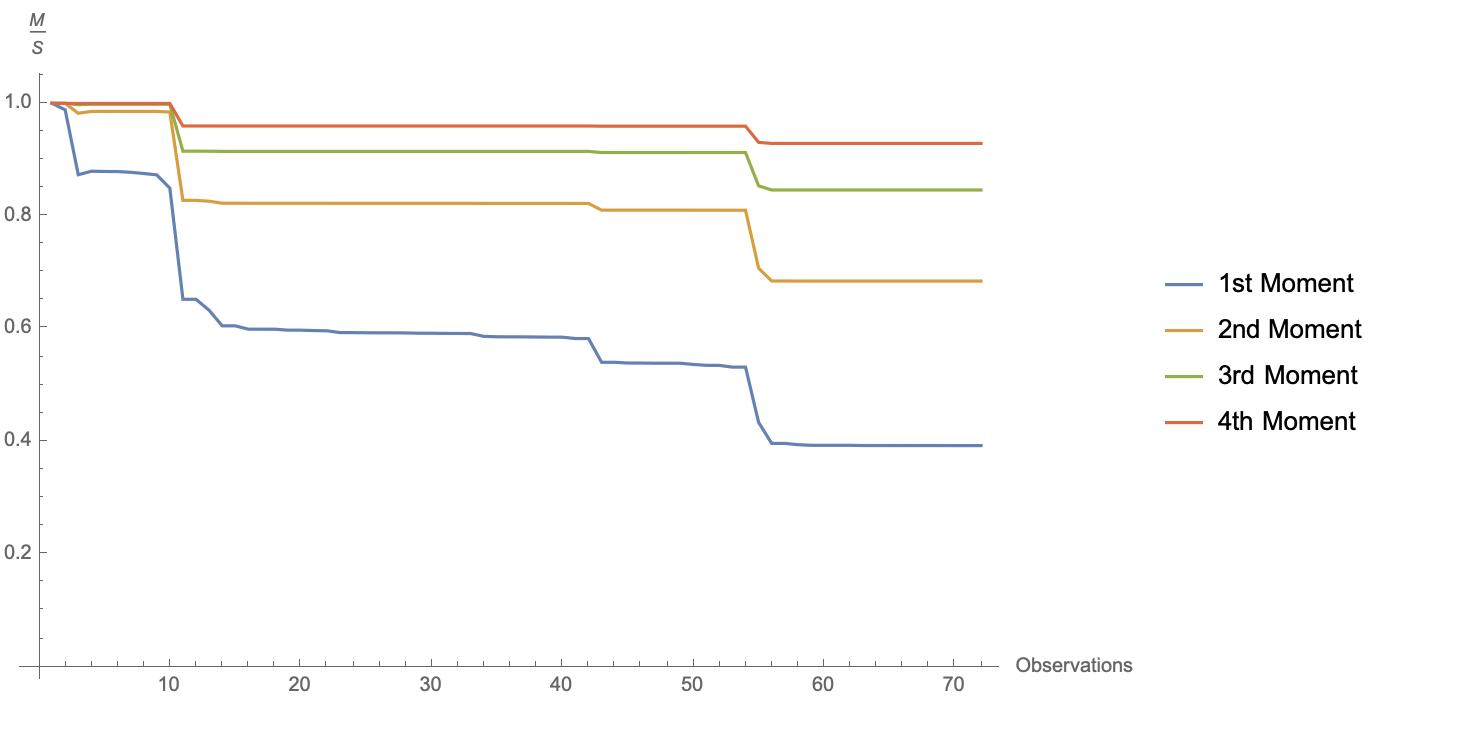}
\caption{Maximum to Sum plot (MS plot) of the average death numbers in pandemic events in history, as per Table 1.}
\label{msplot}
\end{figure}

The core of the problem is shown in Figure \ref{msplot}, with the Maximum-to-Sum plot \cite{Embrechts} of the number of pandemic fatalities in history (data in Table 1). Such a plot relies on a simple consequence of the law of large numbers: for a sequence $X_1,X_2,...,X_n$ of nonnegative i.i.d. random variables, if $E[X^p]<\infty$ for $p=1,2,3...$, then $R_n^p=M_n^p / S_n^p \to^{a.s.} 0$ as $n \to \infty$, where $S_n^p=\sum_{i=1}^n X_i^p$ is the partial sum of order $p$, and $M_n^p=\max(X_1^p,...,X_n^p)$ the corresponding partial maximum. Figure \ref{msplot} clearly shows that no finite moment is likely to exist for the number of victims in pandemics, as the $R_n$ ratio does not converge to 0 for $p=1,2,3,4$, no matter how many data points we use. Such a behavior hints that the victims distribution has such a fat right tail that not even the first theoretical moment is finite. We are looking at a phenomenon for which observed quantities such as the naive sample average and  standard deviation are therefore meaningless for inference.

However, Figure \ref{msplot} (or a naive use of EVT) does not imply that pandemic risk is actually infinite and there is nothing we can do or model. Using the methodology we developed to study war casualties \cite{CiTa,TaCi}, we are in fact able to extract useful information from the data, quantifying the large yet finite risk of pandemic diseases. The method provides in fact rough estimates for quantities not immediately observable in the data.

\subsection*{The tail wags the dog effect}
Centrally, the more fat-tailed the distribution, the more ``the tail wags the dog", that is, the more statistical information resides in the extremes and the less in the ``bulk" (that is the events of high frequency), where it becomes almost noise. This makes EVT the most effective approach, and our sample of extremes very highly sufficient and informative for risk management purposes\footnote{Since the law of large numbers works slowly under fat tails, the bulk becomes increasingly dominated by noise, and averages and higher moments--even when they exist--become uninformative and unreliable, while extremes are rich in information \cite{TaStat}.}.

The fat-tailedness of the distribution of pandemic fatalities has the following policy implications, useful in the wake of the Covid-19 pandemic. 

First, it should be evident that one cannot compare fatalities from multiplicative infectious diseases (fat-tailed, like a Pareto) to those from car accidents, heart attacks or falls from ladders (thin-tailed, like a Gaussian). Yet this is a common (and costly) error in policy making, and in both the decision science and the journalistic literature\footnote{Sadly, this mistake is sometimes made by professional statisticians as well. Thin tailed (discrete) variables are subjected to Chernov bounds, unlike fat-tailed ones \cite{TaStat}.}. Some research papers even criticise people's ``parano{\"i}a" with respect to pandemics, not understanding that such a parano{\"i}a is merely responsible (and realistic) risk management in front of potentially destructive events \cite{TaStat}. The main problem is that those articles--often relied upon for policy making --consistently use the wrong thin-tailed distributions, underestimating tail risk, so that every conservative or preventative reaction is bound to be considered an overreaction.

Second, epidemiological models like the SIR \cite{He} differential equations, sometimes supplemented with simulative experiments like \cite{Ferguson}, while useful for scientific discussions for the bulk of the distributions of infections and deaths, or to understand the dynamics of events after they happened, should never be used for precautionary risk management
, which should focus on maxima and tail exposures instead. It is highly unrigorous to use naive (and reassuring) statistics, like the expected average outcome of compartmental models, or one or more point estimates, as a motivation for policies. Owing to the compounding effect of parameters' uncertainty, the ``tail wagging the dog" effect easily invalidates both point estimates and scenario analyses\footnote{The current Covid-19 pandemic is generating a lot of research, and finally some scholars are looking at the impact of parameters' uncertainty on the scenarios generated by epidemiological models, e.g. \cite{Donnat}.}.

EVT is the natural candidate to handle pandemics. It was born to cope with maxima \cite{Falk}, and it evolved to deal with tail risk in a robust way, even with a limited number of observations and the uncertainty associated with it \cite{Embrechts}. In the Netherlands, for example, EVT was used to get a handle on the distribution of the maxima--not the average!--of sea levels in order to build dams and dykes high and strong enough for the safety of citizens \cite{deHaan}. 

Finally, EVT-based risk management is compatible with the (non-na{\"i}ve) precautionary principle of \cite{Prec}, which should be the leading driver for policy decisions under jointly systemic and extreme risks.

\section{Data and descriptive statistics}

We investigate the distribution of deaths from the major epidemic and pandemic diseases of history, from 429 BC until now. The data are available in Table 1, together with their sources, and only refer to events with more than 1K estimated victims, for a total of 72 observations. As a consequence, potentially high-risk diseases, like the Middle East Respiratory Syndrome (MERS), do not appear in our collection\footnote{Up to the present, MERS has killed 858 people as reported in \url{https://www.who.int/emergencies/mers-cov/en}. For SARS the death toll is between 774 and 916 victims until now \url{https://www.nytimes.com/2003/10/05/world/taiwan-revises-data-on-sars-total-toll-drops.html}.}. All diseases whose end year is 2020 are to be taken as still occurring worldwide, as for the running COVID-19 pandemic.

Three estimates of the reported cumulative death toll have been used:  minimum,  average and maximum. When the three numbers coincide in Table 1, our sources simply do not provide intervals for the estimates. Since we are well aware of the volatility and possible unreliability of historical data \cite{Seybolt,TaCi}, in Section \ref{reliability} we deal with such an issue by perturbing and omitting observations.

In order to compare fatalities with respect to the coeval population (that is, the relative impact of pandemics),  column \textit{Rescaled} of Table 1 provides the rescaled version of column \textit{Avg Est}, using the information in column \textit{Population}\footnote{Population estimates are by definitions estimates, and different sources can give different results (most of the times differences are minor), especially for the past. However our methodology is robust to this type of variability, as we stress later in the paper.} \cite{Goldewijk,Klein,UN}. For example, the Antonine plague of 165-180 killed an average of 7.5M people, that is to say 3.7\% of the coeval world population of 202M people. Using today's population, such a number would correspond to about 283M deaths, a terrible hecatomb, killing more people than WW2.

For space considerations, 
we restrict our attention to the actual average estimates in Table 1, but all our findings and conclusions hold true for the lower, the upper and the rescaled estimates as well\footnote{The differences in the estimates do not change the main message: we are dealing with an extremely erratic phenomenon, characterised by very fat tails.}.

\begin{figure} 
\centering
\includegraphics[width=\linewidth]{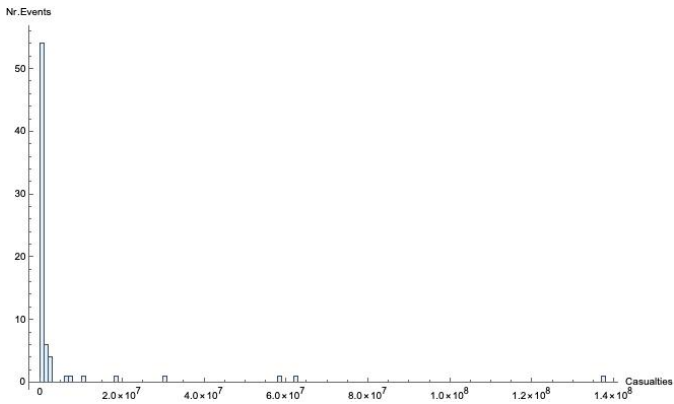}
\caption{Histogram of the average number of deaths in the 72 contagious diseases of Table 1.}
\label{actual}
\end{figure}

Figure \ref{actual} shows the histogram of the actual average numbers of deaths in the 72 large contagious events. The distributions appears highly skewed and possibly fat-tailed. The numbers are as follows: the sample average is 4.9M, while the median is 76K, compatibly with the skewness observable in Figure \ref{actual}. The 90\% quantile is 6.5M and the 99\% quantile is 137.5M. The sample standard deviation is 19M.  

\begin{figure} 
\centering
\includegraphics[width=\linewidth]{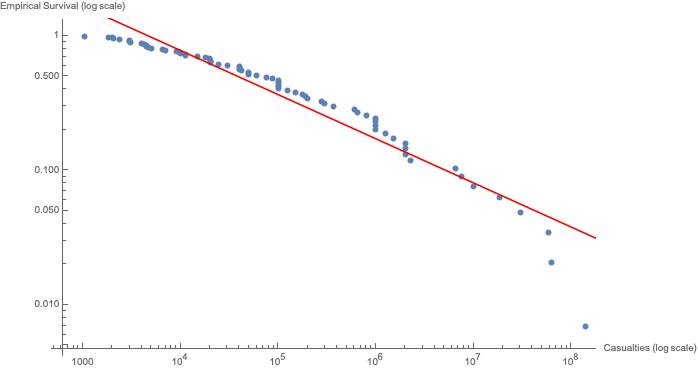}
\caption{Log log plot of the empirical survival function (Zipf plot) of the actual average death numbers in Table 1. The red line represents a naive linear fit of the decaying tail.}
\label{emplot}
\end{figure}

Using common graphical tools for fat tails \cite{Embrechts}, in Figure \ref{emplot} we show the log log plot (also known as Zipf plot) of the empirical survival functions for the average victims over the diverse contagious events. In such a plot possible fat tails can be identified in the presence of a linearly decreasing behavior of the plotted curve. To improve interpretability a naive linear fit is also proposed. Figure \ref{emplot} suggests the presence of fat tails.

\begin{figure} 
\centering
\includegraphics[width=\linewidth]{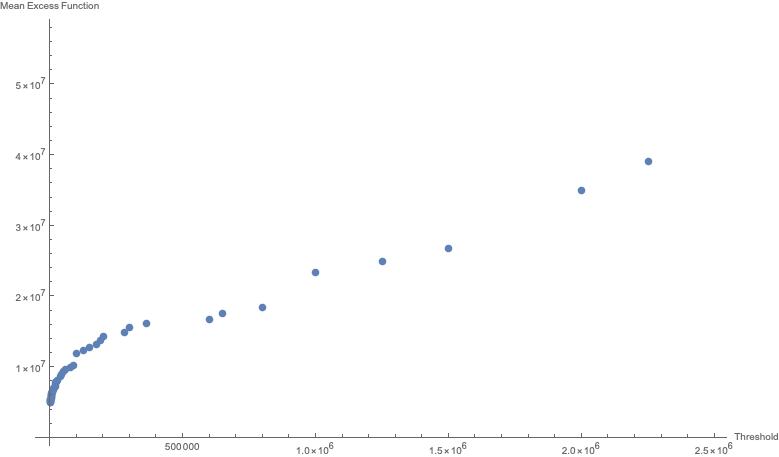}
\caption{Mean excess function plot (meplot) of the average death numbers in Table 1. The plot excludes 3 points on the top right corner, consistently with the suggestions in \cite{Embrechts} about the exclusion of the more volatile observations.}
\label{meplot}
\end{figure}

The Zipf plot shows a necessary but not sufficient condition for fat-tails \cite{Cirillo}. Therefore, in Figure \ref{meplot} we complement the analysis with a mean excess function plot, or meplot. If a random variable $X$ is possibly fat-tailed, its mean excess function $e_X(u)=E[X-u|X\geq u]$ should grow linearly in the threshold $u$, at least above a certain value identifying the actual power law tail \cite{Embrechts}. In a meplot, where the empirical $e_X(u)$ is plotted against the different values of $u$, one thus looks for some (more or less) linearly increasing trend, as the one we observe in Figure \ref{meplot}.

\begin{figure} 
\centering
\includegraphics[width=\linewidth]{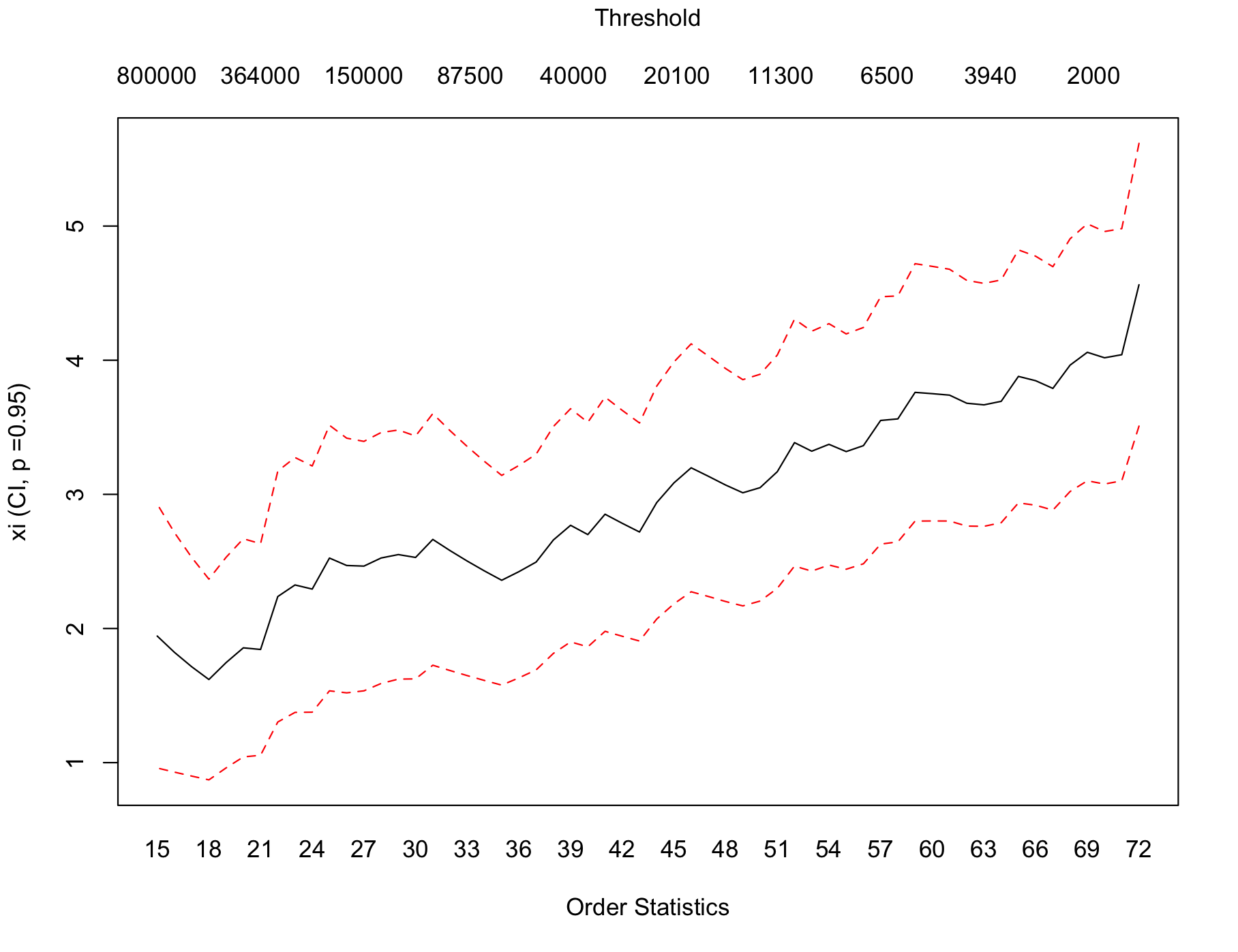}
\caption{Hill plot of the average death numbers in Table 1, with 95\% confidence intervals. Clearly $\xi>1$, suggesting the non-existence of moments.}
\label{hill}
\end{figure}

A useful tool for the analysis of tails--when one suspects them to be fat--is the nonparametric Hill estimator \cite{Embrechts}. For a collection $X_1,...,X_n$, let $X_{n,n}\leq ... \leq X_{1,n}$ be the corresponding order statistics. Then we can estimate the tail parameter $\xi$ as 
$$
\hat{\xi}= \frac{1}{k}\sum_{i=1}^k \log (X_{i,n})-\log(X_{k,n}), \qquad 2 \leq k \leq n.
$$
In Figure \ref{hill}, $\hat{\xi}$ is plotted against different values of $k$, creating the so-called Hill plot \cite{Embrechts}. The plot suggests $\xi>1$, in line with Figure \ref{msplot}, further supporting the evidence of infinite moments.

Other graphical tools could be used and they would all confirm the point: we are in the presence of fat tails in the distribution of the victims of pandemic diseases. Even more, a distribution with possibly no finite moment.
 
\subsection*{The dual distribution}  \label{shadow}

As we observed for war casualties \cite{CiTa}, the non-existence of moments for the distribution of pandemic victims is questionable. Since the distribution of victims is naturally bounded by the coeval world population, no disease can kill more people than those living on the planet at a given time time. We are indeed looking at an \textit{apparently} infinite-mean phenomenon, like in the case of war casualties \cite{CiTa,TaCi} and operational risk \cite{CiTaQF}.

Let $[L,H]$ be the support of the distribution of pandemic victims today, with $L>>0$ to ignore small events not officially definable as pandemic \cite{who}. For what concerns $H$, its value cannot be larger than the world population, i.e. 7.7 billion people in 2020\footnote{Today's world population \cite{UN}  can be safely taken as the upper bound also for the past.}. Evidently $H$ is so large that the probability of observing values in its vicinity is in practice zero, and one always finds observations below a given $M<<H<\infty$ (something like 150M deaths using actual data). Thus one could be fooled by data into ignoring $H$ and taking it as infinite, up to the point of believing in an infinite mean phenomenon, as Figure \ref{msplot} suggests. However notice that a finite upper bound $H$--no matter how large it is--is not compatible with infinite moments, hence Figure \ref{msplot} risks to be dangerously misleading.

\begin{figure}[h]
\includegraphics[width=\linewidth]{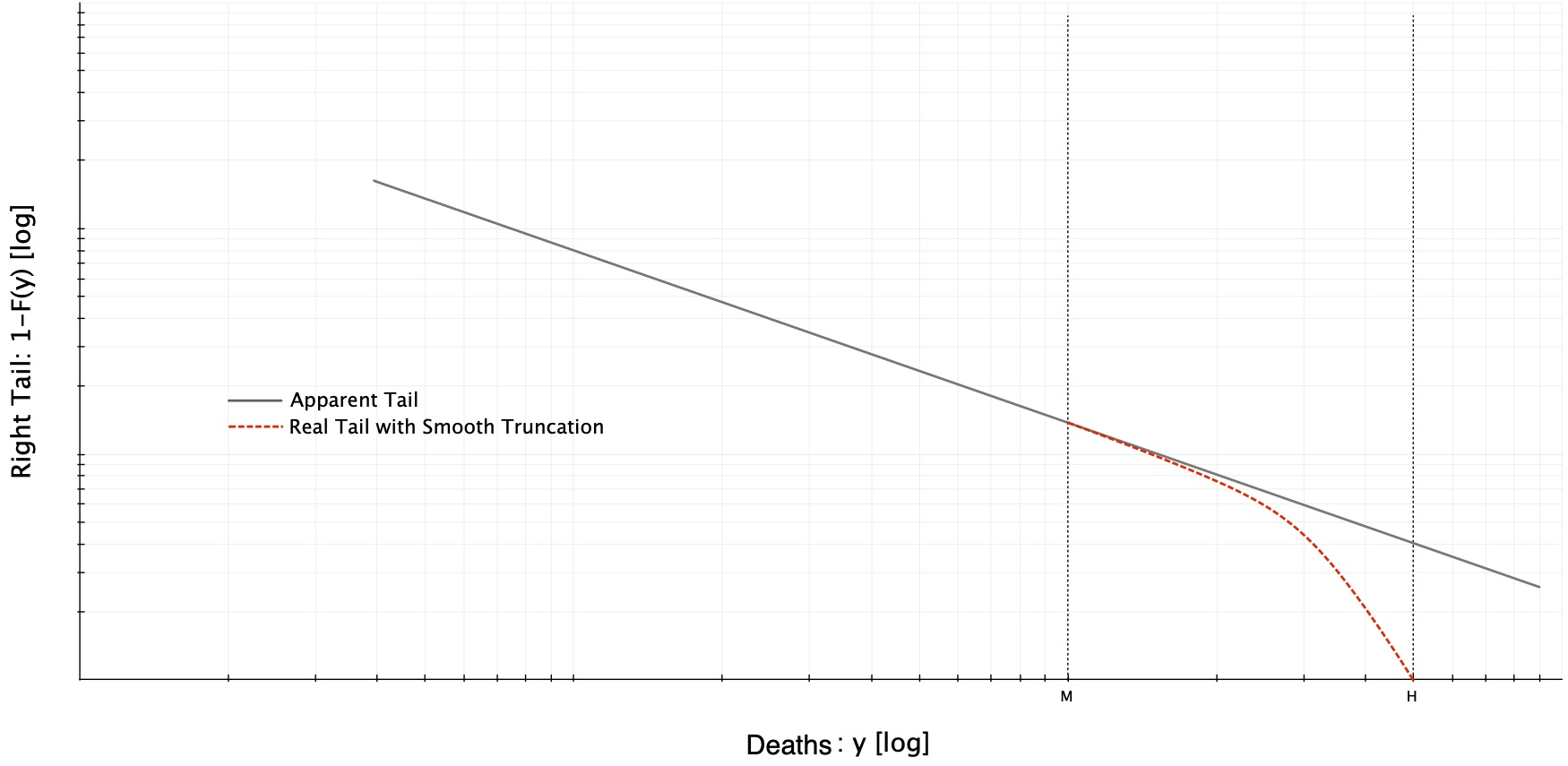}
\caption{Graphical representation (log-log plot) of what may happen if one ignores the existence of the finite upper bound $H$, since only $M$ is observed.}
\label{faketail}
\end{figure}

In Figure \ref{faketail}, the real tail of the random variable $Y$ with remote upper bound $H$ is represented by the dashed line. If one only observes values up to $M<<H$, and more or less consciously ignores the existence of $H$, one could be fooled by the data into believing that the tail is actually the continuous one, the so-called apparent tail \cite{CiTaQF}. The tails are indeed indistinguishable for most cases, virtually in all finite samples, as the divergence is only clear in the vicinity of $H$. A bounded tail with very large upper limit is therefore mistakenly taken for an unbounded one, and no model will be able to see the difference, even if epistemologically we are in two extremely different situations. This is the typical case in which critical reasoning, and the a priori analysis of the characteristics of the phenomenon under scrutiny, should precede any instinctive and uncritical fitting of the data.

A solution is the approach of \cite{CiTa,CiTaQF}, which introduces the concept of dual data via a special log-transformation \footnote{Other log-transformations have been proposed in the literature, but they are all meant to thin the tails, without actually taking care of the upper bound problem: the number of victims can still be infinite. The rationale behind those transformations is given by the observation that if X is a random variable whose distribution function is in the domain of attraction of a Fréchet, the family of fat-tailed distributions, then $\log(X)$ is in the domain of attraction of a Gumbel, the more reassuring family of normals and exponentials \cite{Embrechts}.}. The basic idea is to find a way of matching naive extrapolations (apparently infinite moments) with correct modelling.

Let $L$ and $H$ be respectively the finite lower and upper bounds of a random variable $Y$, and define the function
\begin{equation}
\varphi(Y)= 
L-H \log \left(\frac{H-Y}{H-L}\right).
\label{transform}
\end{equation}
We can easily check that
\begin{enumerate}
\item $\varphi \in C^\infty$,
\item $\varphi^{-1}(\infty)=H,$
\item $\varphi^{-1}(L)=\varphi(L)=L$.
 \end{enumerate} 
Then $Z=\varphi(Y)$ defines a new random variable with lower bound $L$ and an infinite upper bound. Notice that the transformation induced by $\varphi(\cdot)$ does not depend on any of the parameters of the distribution of $Y$, and that $\varphi(\cdot)$ is monotone. From now on, we call the distributions of $Y$ and $Z$, respectively the real and the dual distribution. It is easy to verify that for values smaller than $M<<H$, $Y$ and $Z$ are in practice undistinguishable (and do are their quantiles \cite{CiTaQF}).

As per \cite{CiTa,CiTaQF}, we take the observations in the column "Avg Est" of Table 1, our $Y$'s, and transform them into their dual $Z$'s. We then study the actually unbounded duals using EVT (see Section \ref{tail}), to find out that the naive observation of infinite moments can makes sense in such a framework (but not for the bounded world population!). Finally, by reverting to the real distribution, we compute the so-called \textit{shadow} means \cite{CiTaQF} of pandemics, equal to 
\begin{equation}\label{meanY}
E[Y]=(H-L) e^{\frac{\frac{1}{\xi}   \sigma }{H}} \left(\frac{\sigma }{H \xi }\right)^{\frac{1}{\xi}  } \Gamma \left(1-\frac{1}{\xi}  ,\frac{\sigma }{H \xi}\right)+L,	
\end{equation}
where $\Gamma(\cdot,\cdot)$ is the gamma function.

Notice that the random quantity $Y$ is defined above $L$, therefore its expectation corresponds to a tail expectation with respect to the random variable $Z$, an expected shortfall in the financial jargon, being only valid in the tail above $\mu$ \cite{CiTa}. All moments of the random variable $Y$ are called shadow moments in \cite{CiTaQF}, as they are not immediately visible from the data, but from plug-in estimation.

\section{The dual tail via EVT and the shadow mean} \label{tail}

Take the dual random variable $Z$ whose distribution function $G$ is unknown, and let $z_G=\sup \{z \in \mathbb{R}:G(z)<1\}$ be its right-end point, which can be finite or infinite. Given a threshold $u<z_G$, we can define the exceedance distribution of $Z$ as
\begin{equation} \label{exceedance}
G_u(z) = P(Z \leq z | Z>u) = \frac{G(z)-G(u)}{1-G(u)}, 
\end{equation}
for $z \geq u$.

For a large class of distributions $G$, and high thresholds $u \rightarrow z_G$, $G_u$ can be approximated by a Generalized Pareto distribution (GPD) \cite{deHaan}, i.e. 
\begin{equation} 
G_u(z) \approx GPD(z; \xi, \beta, u)=\begin{cases}
1-(1+\xi \frac{z-u}{\beta})^{-1/ \xi} & \xi \neq 0 \\
1-e^{-\frac{z-u}{\beta}} & \xi = 0 
\end{cases},
\label{GPD}
\end{equation}
where $z\geq u$ for $\xi \geq 0$, $u \leq z \leq u - \beta/\xi$ for $\xi<0$, $u \in \mathbb{R}$, $\xi \in \mathbb{R}$ and $\beta>0$. 

Let us just consider $\xi>0$, being $\xi=0$ not relevant for fat tails. From equation \eqref{exceedance}, we see that $G(z)=(1-G(u))G_u(z)+G(u)$, hence we obtain 
$$G(z)\approx(1-G(u))GPD(z; \xi, \beta, u)+G(u)$$
$$=1-\bar{G}(u)\left(1+\xi \frac{z-u}{\beta}\right)^{-1/\xi},$$ 
with $\bar{G}(x)=1-G(x)$. The tail of $Z$ is therefore
\begin{equation} \label{tail}
\bar{G}(z)=\bar{G}(u)\left(1+\xi \frac{z-u}{\beta}\right)^{-1/\xi}.
\end{equation}

Equation \eqref{tail} is called the tail estimator of $G(z)$ for $z \geq  u$. Given that $G$ is in principle unknown, one usually substitutes $G(u)$ with its empirical estimator $n_u/n$, where $n$ is the total number of observations in the sample, and $n_u$ is the number of exceedances above $u$.

Equation (\ref{tail}) then changes into
\begin{equation}  \label{fullGPD}
\bar{G}(z)=\frac{n_u}{n}\left(1+\xi \frac{z-u}{\beta}\right)^{-1/\xi} \approx 1-GPD(z^\ast;\xi,\sigma,\mu),
\end{equation}
where $\sigma=\beta\left(\frac{n_u}{n} \right)^\xi$, $\mu=u-\frac{\beta}{\xi}\left(1-\left(\frac{n_u}{n} \right)^\xi \right)$, and $z^\ast\geq \mu$ is an auxiliary random variable. Both $\sigma$ and $\mu$ can be estimated semi-parametrically, starting from the estimates of $\xi$ and $\beta$ in equation \eqref{GPD}. If $\xi>-1/2$, the preferred estimation method is maximum likelihood \cite{deHaan}, while for $\xi \leq -1/2$ other approaches are better used \cite{Embrechts}. For both the exceedances distribution and the recovered tail, the parameter $\xi$ is the same, and it also coincides with the tail parameter we have used to define fat tails\footnote{Moreover, when maximum likelihood is used, the estimate of $\xi$ would correspond to $1/\alpha$, where $\alpha$ is estimated according to \cite{Clauset}.}.

One can thus study the tail of $Z$ without caring too much about the rest of the distribution, i.e. the part below $u$. All in all, the most destructive risks come from the right tail, and not from the first quantiles or even the bulk of the distribution. The identification of the correct $u$ is a relevant question in extreme value statistics \cite{deHaan, Embrechts}. One can rely on heuristic graphical tools \cite{Cirillo}, like the Zipf plot and the meplot we have seen before, or on statistical tests for extreme value conditions \cite{Falk} and GPD goodness-of-fit \cite{Arshad}. 

What is important to stress--once again--is that the GPD fit needs to be performed on the dual quantities, to be statistically and epistemologically correct. One could in fact work with the raw observation directly, without the log-transformation of Equation \eqref{transform}, surely ending up with $\xi>1$, in line with Figures \ref{msplot} and \ref{hill}. But a similar approach would be wrong and naive, because only the dual observations are actually unbounded.

Working with the dual observations, we find out that the best GPD fit threshold is around 200K victims, with 34.7\% of the observations lying above. For what concerns the GPD parameters, we estimate $\xi=1.62$ (standard error $0.52$), and $\beta=1'174.7K$ (standard error $536.5K$). As expected $\xi>1$ once again supporting the idea of an infinite first moment\footnote{Looking at the standard error of $\xi$, one could argue that, with more data from the upper tail, the first moment could possibly become finite, yet there would be no discussion about the non existence of the second moment, and thus the unreliability of the sample mean \cite{TaStat}. Pandemic fatalities would still be an extremely erratic phenomenon, with substantial tail risk in the number of fatalities. In any case, Figures \ref{msplot} and \ref{hill} make us prefer to consider the first moment as infinite, and not to trust sample averages.}. Visual inspections and statistical tests \cite{Arshad,Falk} support the goodness-of-fit for the exceedance distribution and the tail.

Given $\xi$ and $\beta$, we can use Equations \eqref{meanY} and \eqref{fullGPD} to compute the shadow mean of the numbers of victims in pandemics. For actual data we get a shadow mean of 20.1M, which is definitely larger (almost 1.5 times) than the corresponding sample tail mean of 13.9M  (this is the mean of all the actual numbers above the 200K threshold.). Combining the shadow mean with the sample mean below the $200K$ threshold, we get an overall mean of 7M instead of the naive 4.9M we have computed initially. It is therefore important to stress that a naive use of the sample mean would induce an underestimation of risk, and would also be statistically incorrect.

\section{Data reliability issues} \label{reliability}

As observed in \cite{CiTa,Seybolt,TaCi} for war casualties, but the same reasoning applies to pandemics of the past, the estimates of the number of victims are not at all unique and precise. Figures are very often anecdotal, based on citations and vague reports, and usually dependent on the source of the estimate. In Table 1, it is evident that some events vary considerably in estimates.

Natural questions thus arise: are the tail risk estimates of Section \ref{tail} robust? What happens if some of the casualties estimates change? What is the impact of ignoring some events in our collection? The use of extreme value statistics in studying tail risk already guarantees the robustness of our estimates to changes in the underlying data, when these lie below the threshold $u$. However, to verify robustness more rigorously and thoroughly, we have decided to stress the data, to study how the tails potentially vary.

First of all, we have generated 10K distorted copies of our dual data. Each copy contains exactly the same number of observations as per Table 1, but every data point has been allowed to vary between $80\%$ and $120\%$ of its recorded value before imposing the log-transformation of Equation \eqref{transform}. In other words, each of the 10K new samples contains 72 observations, and each observation is a (dual) perturbation ($\pm 20\%$) of the corresponding observation in Table 1.

\begin{figure} 
\centering
\includegraphics[width=\linewidth]{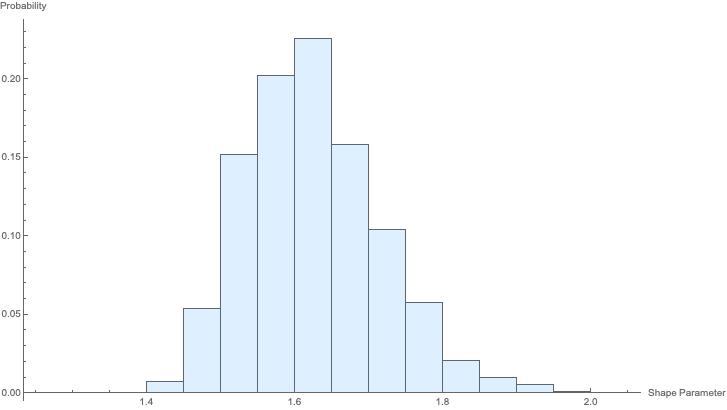}
\caption{Values of the shape parameter $\xi$ over 10,000 distorted copies of the the dual versions of the average deaths in Table 1, allowing for a random variation of $\pm 20\%$ for each single observation. The $\xi$ parameter consistently indicates an apparently infinite-mean phenomenon.}
\label{distorted}
\end{figure}

Figure \ref{distorted} contains the histogram of the $\xi$ parameter over the 10K distorted copies of the dual numbers. The values are always above 1, indicating an apparently infinite mean, and the average value is $1.62$ (standard deviation 0.10), in line with our previous findings. Our tail estimates are thus robust to imprecise observations. Consistent results hold for the $\beta$ parameter.

\begin{figure} 
\centering
\includegraphics[width=\linewidth]{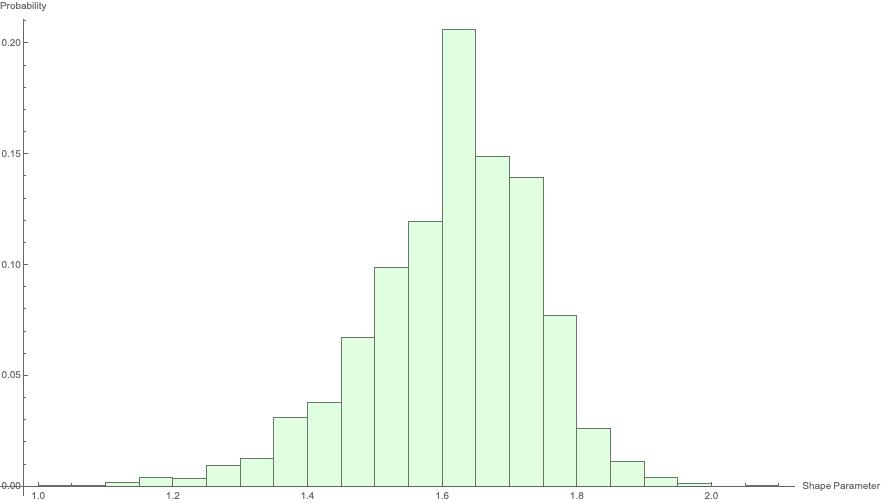}
\caption{Values of the shape parameter $\xi$ over 10,000 jackknifed versions of the dual versions of the actual average numbers in Table 1, when allowing at least 1\% and up to about 10\% of the observations to be missing. The $\xi$ parameter consistently indicates an apparently infinite-mean phenomenon.}
\label{distortedjk}
\end{figure}

But it also true that our data set is likely to be incomplete, not containing all epidemics and pandemics with more than 1K victims, or that some of the events we have collected are too biased to be reliable and should be discarded anyway. To account for this, we have once again generated 10K copies of our sample via jackknife. Each new dual sample is obtained by removing from 1 to 7 observations at random, so that one sample could not contain the Spanish flu, while another could ignore the Yellow Fever and AIDS. In Figure \ref{distortedjk} we show the impact of such a procedure on the $\xi$ parameter. Once again, the main message of this work remains unchanged: we are looking at a very fat-tailed phenomenon, with an extremely large tail risk and potentially destructive consequences, which should not be downplayed in any serious policy discussion.

\begin{figure*}
\centering
\includegraphics[scale=0.25]{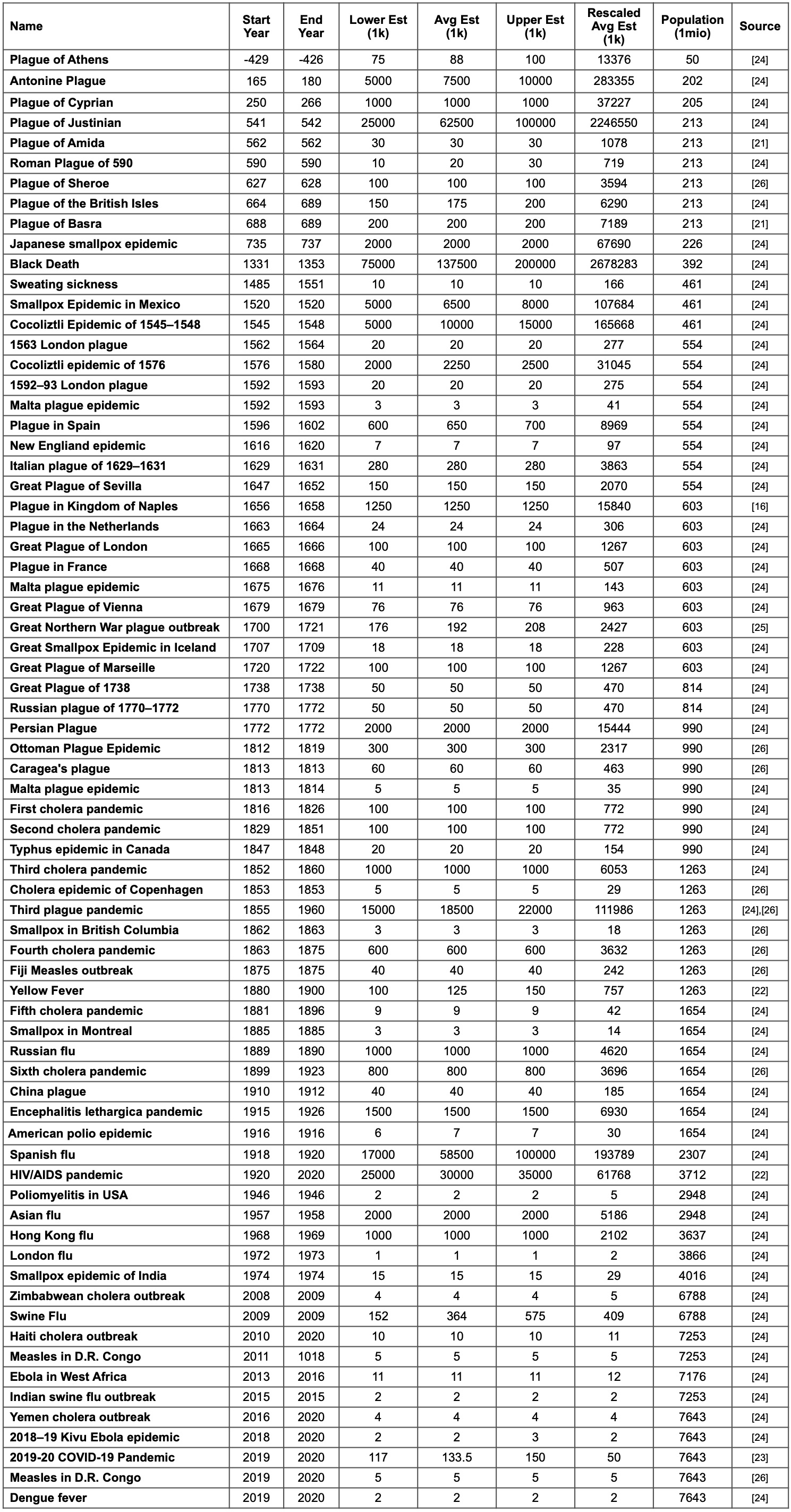}
\caption*{Table 1: The data set used for the analysis. All estimates in thousands, apart from coeval population, which is expressed in millions. For Covid-19 \cite{who}, the upper estimate includes the supposed number of Chinese victims (42K) for some Western media.}
\label{data}
\end{figure*}

\end{document}